\documentclass[aps,twocolumn,raggedbottom,prl,showpacs,nobalancelastpage,amssymb,groupedaddress]{revtex4}
\usepackage{graphicx}
\usepackage{amsmath}
\usepackage{psfrag}

\begin{document}
\title{Two-particle scattering matrix of two interacting mesoscopic conductors}
\author{M. C.  Goorden and M.\  B\"{u}ttiker}
\affiliation{D\'epartement de Physique Th\'eorique, Universit\'e de Gen\`eve,
  CH-1211 Gen\`eve 4, Switzerland.}
\date{Oct. 5 2007}
\begin{abstract}
We consider two quantum coherent conductors interacting weakly via long range Coulomb forces. We describe the interaction in terms of two-particle collisions described by a two-particle scattering matrix. As an example we determine the
transmission probability and correlations in a two-particle scattering experiment and find
that the results can be expressed in terms of the density-of-states matrices of
the non-interacting scatterers.
\end{abstract}
 \maketitle

Recently there has been a growing interest in intriguing subjects such as quantum measurement, controlled dephasing and shot noise correlations which involve two or more separate mesoscopic conductors interacting via long range Coulomb forces. We highlight here only three recent experiments. The noise cross-correlation of two
capacitively coupled quantum dots has been measured \cite{McC07} and the sign of this
correlation was found to be tunable by the gate voltage. In another work
a Mach-Zehnder interferometer \cite{Ji03,litv,roche} coupled to a detector
channel with shot noise was investigated both experimentally and theoretically
\cite{NedPRL,Ned07}. 
Ref.\ \cite{Suk07} shows how the current through a quantum point contact,
capacitively coupled to a quantum dot, can be used to determine the statistics of charge transferred through the quantum dot. 

Several approaches have been developed to treat Coulomb coupled conductors. For systems in the tunneling limit, a particle number resolved master equation approach \cite{Gur97,Kie06} is often used. For systems well connected to contacts a self-consistent scattering approach
has been developed and has been applied to dynamic conductance and charge fluctuations \cite{Ped98,But99,hekk} in good agreement with experiment \cite{gabe}. A momentum resolved treatment of long range Coulomb interaction is  necessary to treat the Coulomb drag which one conductor exerts on another \cite{Mor01,Yama}. 

We are interested in a different formulation based on two-particle collision processes. We view two-particle interactions as the elementary process and derive a two-particle scattering matrix. In the weak interaction limit such an approach can be expected to have a wide range of applicability similar to Boltzmann equations with two-particle collision kernels. In contrast, in the strong interaction limit, it will be necessary to go beyond two-particle processes and permit for instance the interaction of a carrier in one conductor with a number of carriers in the other conductor \cite{Ned07}.

Two-particle processes occur also if both particles are in the same conductor. Even in the absence of interactions, shot noise tests two-particle correlations and a predicted two-particle Aharonov-Bohm effect \cite{Sam04} has recently been measured \cite{Ned07exp}.
With interaction a number of highly interesting effects have been discussed in
systems with disorder \cite{Jac97}, for conductance and pumping \cite{Sun03} and noise \cite{Sel06} in quantum dots.
In distinction, we extend scattering theory and obtain results that are both very general yet have still an immediate physical appeal.  

Interestingly the description of two-particle processes presented here involves, like the self-consistent scattering approach, a generalization of the Wigner-Smith delay time matrix \cite{Smi60,Bro97}. 
The Wigner-Smith matrix contains energy derivatives of the scattering matrix: its diagonal elements are proportional to the density of states generated by scattering states incident from a particular contact \cite{Ped98}. Thus diagonal elements are useful to describe the piled up charges in a conductor in response to a voltage change at a contact. The off-diagonal elements of the matrix describe spontaneous charge fluctuations \cite{Ped98,But99}.

{\begin{figure}[t]
\begin{center}
\psfrag{ald}{$a_L^\dag$}
\psfrag{bld}{$b_L^\dag$}
\psfrag{dld}{$d_L^\dag$}
\psfrag{cld}{$c_L^\dag$}
\psfrag{ard}{$a_R^\dag$}
\psfrag{brd}{$b_R^\dag$}
\psfrag{drd}{$d_R^\dag$}
\psfrag{crd}{$c_R^\dag$}
\psfrag{1}{$\hat{Q}^{\rm I}$}
\psfrag{2}{$\hat{Q}^{\rm II}$}
\psfrag{l12}{$\lambda \hat{Q}^{\rm I}\hat{Q}^{\rm II}$}
\includegraphics[width=5cm]{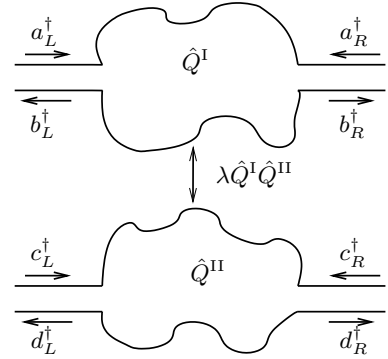}
\end{center}
\caption{Two quantum dots coupled via the interaction
$\lambda\hat{Q}^{\rm I}\hat{Q}^{\rm II}$. The operators $a^\dag$ ($c^\dag$) create
incoming electrons in the scattering states in the left (L) and right (R) leads of dot ${\rm I}$ (${\rm II}$), while
  $b^\dag$ ($d^\dag$) are similar operators for outgoing electrons.}
\label{dotsfig}
\end{figure}}

In Fig.\ \ref{dotsfig} we show the type of system we are interested in. 
We consider two scatterers, both coupled to two
non-interacting leads $i = L, R$. For simplicity we consider single channel leads. 
In the figure the scatterers are chaotic quantum dots
(and we will refer to them as such), 
but our theory is valid for any scatterer.
The Hamiltonians of the two (non-interacting) sub-systems read
\begin{equation}
H^{\rm I}=H_{d}^{\rm I}+H_{l}^{\rm I}+V^{{\rm I}},  \hspace{0.4cm} H^{\rm II}=H_{d}^{{\rm II}}+H_{l}^{{\rm II}}+V^{{\rm II}}.
\end{equation}
The symbols ${\rm I}$/${\rm II}$ refer to the first/second conductor.
The lead and dot Hamiltonians are given by
\begin{eqnarray}
H_{l}^{\rm I}&=&\sum_i\int dE E a_i^{\dag}(E)a_i(E),\hspace{0.2cm} H_{d}^{\rm I}=\sum_i f_i^\dag f_i \epsilon_i,\\
H_{l}^{{\rm II}}&=&\sum_i\int dE E c_i^{\dag}(E)c_i(E),\hspace{0.2cm}H_{d}^{{\rm II}}=\sum_i g_i^\dag g_i E_i.
\end{eqnarray}
While  $a^\dag_i(E)$ creates incoming carriers in lead $i$ of the first dot,
$c^\dag_i(E)$ is a similar operator for the second dot.
Furthermore $f_i$ ($g_i$) annihilates an electron with energy $\epsilon_i$
($E_i$) in the first (second) dot.
The coupling between leads and dots is described by
\begin{eqnarray}
V^{\rm I}&=&\sum_{il}\int dE (a^\dag_i(E)f_l W_{il}^{\rm I}+H.c.),\\
V^{{\rm II}}&=&\sum_{il}\int dE (c^\dag_i(E)g_l W_{il}^{{\rm II}}+H.c.).
\end{eqnarray}
We will assume an interaction of the form
\begin{eqnarray}
H_c&=&\frac{\lambda}{e^2}\hat{Q}^{\rm I}\hat{Q}^{{\rm II}}.
\end{eqnarray}
with $\lambda$ a coupling energy and
$\hat{Q}^{{\rm I}}= e \sum_i f_i^\dag f_i$
($\hat{Q}^{{\rm II}}= e \sum_i g_i^\dag g_i$)
the charge operator on dot ${\rm I}$ (${\rm II}$). 

Before proceeding to treat this interacting problem, we recall properties of a
single non-interacting dot (here dot I). The single-particle scattering matrix
$S^{\rm I}(E)$ relates  operators which annihilate carriers in incoming and outgoing
states, $b_i = \sum_j S^{\rm I}_{ij} a_j$. The charge  fluctuations at frequency $\omega = E -E'$ can be expressed in terms of a density of states matrix \cite{Ped98,But99}
\begin{equation}
\label{dens}
{\cal N}^{\rm I}(E,E')=S^{\rm I \dag}(E)\frac{S^{\rm I}(E)-S^{\rm I}(E')}{2\pi i(E-E')}.
\end{equation}
The diagonal elements of this matrix ${\cal N}^{\rm I}_{LL}$ (and ${\cal
  N}^{\rm I}_{RR}$) are the injectance of the left (right) contact, i.e. the
  part of the total density of states associated with carriers incident in the
  left (right) lead. In the limit $E\rightarrow E'$ the density of states
  matrix reduces to the famous Wigner-Smith delay
time matrix discussed in the introduction. 

We are interested in correlations generated by the Coulomb interaction between the two dots. 
Consider the particles leaving the two dots. Particles in the outgoing states
  are created by operators $b^\dag_i(E)$, $d_i^\dag(E)$. The outgoing two-particle state $b_i^\dag(E_1)d_j^\dag(E_2 )$ depends on two-particle input states in all pairs of incident channels and depends on the energy of the incident carriers not only at energy $E_1$ and $E_2$ but also at $E_1 - \omega $ and $E_2 + \omega$ where $\omega$
is the energy which can be exchanged in the collision process. As a consequence the relation between incoming and outgoing two-particle states is 
\begin{eqnarray}
\label{invsout}
b_i(E_1)d_j(E_2)&=&\sum_{kl}\left[S_{ik}^{\rm I}(E_1)S_{jl}^{\rm
  II}(E_2)a_k(E_1)c_l(E_2)+\right.\nonumber\\
&&\int d\omega\delta S_{ik,jl}(E_1,E_2,E_1-\omega,E_2+\omega)\nonumber\\
&&\left. a_k(E_1-\omega) c_l(E_2+\omega)\right].
\end{eqnarray}
Eq.\ (\ref{invsout}) defines the two-particle scattering matrix.
The first part of the relation is simply a product of the two single-particle scattering matrices, the second part contains the effect of the interaction. In this paper we will
derive $\delta S$ up to first order in the interaction energy
$\lambda$.

Let us first illustrate the derivation of the single dot scattering matrix (we
choose dot ${\rm I}$). 
We denote the $n$'th eigenstate of the Hamiltonian $H_{d}^{\rm I}+H_{l}^{\rm I}$ with energy $E$ by $\phi_{nE}^{\rm I}$.
The eigenstates are either localized in the dot or lead. Since the spectrum in the lead is continuous, matrix multiplication also involves an integral over the energy.
We are interested in the eigenstate $\psi^{{\rm I}}_{nE}$ of $H^{\rm I}$ which approaches $\phi_{nE}^{\rm I}$ in the limit $V^{\rm I}\rightarrow 0$.
It is expressed by the Lippmann-Schwinger equation \cite{Tay72}
\begin{equation}
\label{lippmann}
|\psi^{{\rm I}\pm}_{nE}\rangle=|\phi_{nE}^{\rm I}\rangle+G^{{\rm I}\pm}_EV^{\rm I}|\psi^{{\rm I} \pm}_{nE}\rangle,
\end{equation}
with $G^{{\rm I}\pm}_E=\left({E-H_d^{\rm I}-H_l^{\rm I}\pm i\eta}\right)^{-1}$.
Here $\eta$ is a positive infinitesimal.
It can be shown (see Ref.\ \cite{Tay72}) that $|\psi_{nE}^{{\rm I}+}\rangle$ ($|\psi_{nE}^{{\rm I}-}\rangle$) satisfies the boundary condition for ingoing (outgoing) states.
The scattering matrix relates the two and is therefore
defined to be
$S_{nm}^{\rm I}(E)\delta(E-E')=\langle\psi_{nE}^{{\rm I}-}|\psi_{mE'}^{{\rm I}+}\rangle$.
Using Eq.\ (\ref{lippmann}) one can rewrite it as
\begin{equation}
\label{SvsT}
S_{nm}^{\rm I}(E)=\delta_{nm}-2\pi i {\cal T}_{nm}^{\rm I}(E,E),
\end{equation}
with transition matrix ${\cal T}_{nm}^{\rm I}(E,E')=\langle\phi^{\rm I}_{nE}|V^{\rm
  I}|\psi_{mE'}^{{\rm I}+}\rangle$ \cite{Tay72}.
For later use we will not only be interested in the transition matrix between
different lead states \cite{Hac01},  but also between lead (l) and dot (d)
states. Using the Lippmann-Schwinger equation we find
\begin{eqnarray}
\label{Tll}
{\cal T}^{{\rm I}(ll)}(E,E')&=&W^{{\rm I}}{\left[D^{\rm I}(E')\right]^{-1}}W^{{\rm I}\dag} ,\\
\label{Tld}
{\cal T}^{{\rm I} (ld)}(E,E')&=&W^{{\rm I}}{\left[D^{\rm I}(E')\right]}^{-1}\left(E'-H_d^{\rm I}\right),\\
\label{Tdl}
{\cal T}^{{\rm I}(dl)}(E,E')&=&\left(E'-H_{d}^{{\rm I}}\right){\left[D^{\rm I}(E')\right]}^{-1}W^{{\rm I}\dag},
\end{eqnarray}
with
$D^{\rm I}(E)={E-H_{d}^{{\rm I}}+i\pi W^{{\rm I}\dag} W^{{\rm I}}}$.
Combining Eqs.\  (\ref{SvsT}) and (\ref{Tll}) gives the well-known scattering
matrix relating in- and outgoing lead states \cite{Bee97}
\begin{equation}
S^{{\rm I}}(E)=\openone^{\rm I}-2\pi i W^{{\rm I}}\left[D^{\rm I}(E)\right]^{-1}W^{{\rm I}\dag} .
\end{equation}

We will now turn to the case of two quantum dots. Without interaction
$(\lambda=0)$  the eigenstates are product
states $|\psi^{\rm I}_{nE_1}\rangle\otimes
|\psi_{mE_2}^{{\rm II}}\rangle\equiv|\psi^{\rm I}_{nE_1}\psi_{mE_2}^{{\rm II}}\rangle$. 
Let $|\zeta^\pm_{nE_1,m E_2}\rangle$ be the eigenstate of the Hamiltonian
$H^{\rm I}\otimes\openone^{\rm II}+\openone^{\rm I}\otimes H^{{\rm II}}+H_c$ which approaches $|\psi^{{\rm I}\pm}_{nE_1}\psi_{mE_2}^{{\rm II}\pm}\rangle$
if $\lambda\rightarrow 0$.
It fulfills a modified Lippmann-Schwinger equation
\begin{equation}
\label{lippmann2}
|\zeta^{\pm}_{nE_1,m E_2}\rangle=|\psi_{nE_1}^{{\rm I}\pm}\psi_{m E_2}^{{\rm II}\pm}\rangle+G^\pm_{E_1+E_2}H_c|\zeta^{\pm}_{nE_1,m E_2}\rangle,
\end{equation}
with $G^\pm_E=({E-H^{\rm I}\otimes\openone^{\rm II}-\openone^{\rm I}\otimes H^{{\rm II}}\pm i\eta})^{-1}$.
The two-particle scattering matrix has matrix elements
\begin{eqnarray}
S_{nm,kl}(E_1,E_2,E_3,E_4)\delta(E_1+E_2-E_3-E_4)=\nonumber\\\langle\zeta_{nE_1,k
  E_2}^{-}|\zeta_{mE_3,l E_4}^{+}\rangle.
\end{eqnarray}
With the help of the Lippmann-Schwinger equation, we find the interacting part
\begin{eqnarray}
\label{T2}
&&\hspace{-0.9cm}\delta S_{nm,kl}(E_1,E_2,E_3,E_4)=\nonumber\\
&&\hspace{0.9cm}-2\pi i\langle\psi_{k E_2}^{\rm
  II-}\psi_{n E_1}^{\rm I-}|H_c|\zeta_{mE_3,l  E_4}^+\rangle=\nonumber\\
&&\hspace{0.9cm}-2\pi i\langle\psi_{k E_2}^{\rm II-}\psi_{nE_1}^{\rm I-}|H_c|\psi^{{\rm I}+}_{mE_3}\psi^{{\rm II} +}_{l  E_4}\rangle+O(\lambda^2).
\end{eqnarray}
For the last equality we have used Eq.\ (\ref{lippmann2}) and expanded up to
first order in the coupling energy $\lambda$.
Because the coupling term $H_c$ is a direct product of operators working on
dot ${\rm I}$
and dot ${\rm II}$, we can write 
\begin{eqnarray}
\label{expT2}
\langle\psi_{k E_2}^{\rm II-}\psi^{\rm I-}_{n E_1}|H_c|\psi^{\rm I+}_{mE_3}\psi^{\rm II+}_{l
  E_4}\rangle&=&\frac{\lambda}{e^2}\langle\psi_{n
  E_1}^{{\rm I}-}|\hat{Q}^{\rm I}|\psi^{{\rm I}+}_{mE_3}\rangle\times\nonumber\\&&\langle\psi_{k
  E_2}^{{\rm II}-}|\hat{Q}^{{\rm II}}|\psi^{{\rm II}+}_{l E_4}\rangle.
\end{eqnarray}
We have expressed everything in single dot quantities and we can use the single dot Lippmann-Schwinger equation to proceed. The operator $\hat{Q}^{\rm I}$ is
the charge operator of the {\em dot} and therefore
\begin{equation}
\label{exp2T2}
\langle\psi_{nE_1}^{{\rm I}-}|\hat{Q}^{\rm I}|\psi^{{\rm I}+}_{mE_3}\rangle=e\sum_{k} \langle\psi_{nE_1}^{{\rm I}-}|\phi^{\rm I(d)}_{kE}\rangle\langle\phi^{\rm I(d)}_{kE}|\psi^{{\rm I}+}_{mE_3}\rangle.
\end{equation}
We only sum over states in the dot as indicated by the superscript $(d)$.
Using Eq.\ (\ref{lippmann}) we find
\begin{eqnarray}
\label{exp3T2}
\langle\phi^{\rm I}_{kE}|\psi^{{\rm I}+}_{mE_3}\rangle=\langle\psi^{{\rm I}-}_{kE_3}|\phi^{{\rm I}}_{mE}\rangle
=\delta_{km}+\frac{{\cal T}^{\rm I}_{km}(E,E_3)}{E_3-E+i\eta}.
\end{eqnarray}
Combining Eqs. (\ref{T2}), (\ref{expT2}), (\ref{exp2T2})  and (\ref{exp3T2})
we find that the
two-particle scattering matrix between different lead
states depends on the single-particle ${\cal T}$-matrices between lead and dot. Using Eqs. (\ref{Tld}) and
(\ref{Tdl}) we calculate
\begin{eqnarray}
\label{deltaS}
\delta S(E_1,E_2,E_3,E_4)=-
2\pi i\lambda  \frac{S^{\rm I}(E_1)-S^{\rm I}(E_3)}{2\pi i(E_1-E_3)}\otimes\nonumber\\ \frac{S^{{\rm II}}(E_2)-S^{{\rm II}}(E_4)}{2\pi i(E_2-E_4)}.
\end{eqnarray}
Thus we have expressed the two-particle scattering matrix in terms of the
scattering matrices of the uncoupled dots. Eq.\ (\ref{deltaS}) is the key result of this work. 
Its form reminds us of the density of states matrix Eq.\ (\ref{dens}).

The application of our two-particle scattering matrix to a transport
experiment such as the experiment in Ref.\ \cite{McC07} requires that we take
into account Pauli blocking in the Fermi sea of the leads.
This is not trivial and we will present our solution in a later work \cite{MGMB}.
Here we will illustrate the properties of the two-particle scattering matrix by considering a real two-particle scattering experiment. 
 We suppose our scattering problems to be one-dimensional and denote the coordinates along the left lead of the first (second) dot by $x$ ($y$).  We will assume that at time $t_0$ a wave packet is created in each left lead, at positions $x_0$ and $y_0$. The dots are at positions $x = 0$ and $y = 0$. This means that we can describe the initial state in our system by
\begin{eqnarray}
\label{psi}
|\Psi\rangle&=&\int dEdE'\alpha^{\rm I}(E)\alpha^{\rm II}(E')e^{i(k(E)x_0+k(E')y_0)}\nonumber\\
&&e^{-i(E+E')t_0/\hbar}
{a}^\dag_L(E)c_L^\dag(E')|0\rangle.
\end{eqnarray}
The functions $\alpha^i(E)$ obey $\int dE |\alpha^i(E)|^2=1$,
and $k(E)>0$ is the wave vector corresponding to an energy $E$.

Since we work with wave packets the timing is important: if both wave packets
reach the dots at very different times they cannot interact. 
Under the assumption
that the width of the wave packet $1/\delta k$ is much larger than $v_F
\tau_d$, the dot appears effectively point like and we can factorize the
influence of the interaction into a
contribution from the wave packet overlap and a contribution from
the scattering matrices of the dots. Here $v_F$ is the Fermi velocity and
$\tau_d$ the dwell time in the dot.  
In this limit it is the integral
\begin{eqnarray}
\label{overlapintegral}
{\cal I}&=&\int
dE_1dE_2d\omega\alpha^{\rm I}(E_1)\alpha^{\rm I *}(E_1+\omega)\alpha^{\rm
  II}(E_2)\alpha^{\rm II *}(E_2-\omega)\nonumber\\&&
e^{i(k(E_1+\omega)-k(E_1))x_0+i(k(E_2-\omega)-k(E_2))y_0},
\end{eqnarray}
that quantifies the overlap of the wave packets in the dot.
For a Gaussian distribution $\alpha^{\rm I/II}(E)=e^{-(E-E_F)^2/(4\delta
  E^2)}/(2\pi\delta E^2)^{1/4}$ with $\delta E\ll E_F$ (so that we can
  linearize the wave vectors around the Fermi wave vector $k_F$) we find
${\cal I}=2\sqrt{\pi}\delta E  \exp{(-(x_0-y_0)^2\delta k^2)}$.
The widths in $k$-space and $E$-space are related by $\delta k=\delta
E({m/2E_F\hbar^2})^{1/2}$, with $E_F$ the Fermi energy and $m$ the mass.
There is only an effect of the interaction if the wave packets are
timed to reach the dots at about the same time,
i.e. for $|x_0-y_0|\delta k\ll 1$ (we assumed equal Fermi velocities).
Furthermore ${\cal I}$ vanishes linearly with $\delta E$, because completely
delocalized particles (plane waves) have a vanishing probability to be in the dot.

Let us first calculate the probability that the first particle leaves through
the right lead, regardless of the behaviour of the second particle. We define the operator $\hat{n}^{\rm I}_R=\int dE b_R^\dag(E)b_R(E)$ and we calculate
\begin{eqnarray}
\label{expvalue}
\langle \hat{n}^{\rm I}_R\rangle=\langle \Psi|\hat{n}^{\rm I}_R|\Psi \rangle.
\end{eqnarray}
We rewrite the state (\ref{psi}) in terms of
output operators $b^\dag$ and $d^\dag$, using the inverse of Eq.\
(\ref{invsout}). Eq.\ (\ref{expvalue}) now contains expectation values of the form $\langle
0|d_i(E_1)b_j(E_2)b_R^\dag(E) b_R(E) b_m^\dag(E_3) d_n^\dag(E_4)|0\rangle=\delta_{in}\delta(E_1-E_4)\delta_{jR}\delta(E_2-E)\delta_{Rm}\delta(E-E_3)$.
In the limit $\delta k\gg v_F\tau_d$ (discussed above Eq.\
(\ref{overlapintegral})) the energy dependence of the scattering matrices can
be neglected and we can evaluate all scattering matrices at the Fermi energy. We will therefore suppress the energy argument. We write
\begin{eqnarray}
\label{expvalue2}
\langle\hat{n}^{\rm I}_R\rangle
&=&T^{\rm I}+{\cal I}
\sum_p (S^{{\rm I}*}_{RL}S^{{\rm II}*}_{pL}\delta S_{RL,pL}+ S^{{\rm
    I}}_{RL}S^{{\rm II}}_{pL}\delta S^*_{RL,pL})\nonumber\\
&=&T^{\rm I}-{\cal I}\lambda\frac{\partial T^I}{\partial E} {{\cal N}}_{LL}^{\rm II}. 
\end{eqnarray}
Here $T^{\rm I}=|S^{\rm I}_{LR}|^2$ is the transmission probability of the first scatterer.
Thus due to the interaction the transmission
probability of the first dot depends on the {\it injectance} of the left lead of
the second dot defined in Eq.\ (\ref{dens}).

Secondly we calculate the cross-correlation between particles in the right leads
\begin{eqnarray}
\label{corr1}
&&\langle \delta\hat{n}^{\rm I}_R\delta\hat{n}^{\rm II}_R\rangle
=
\frac{\lambda{\cal I}}{2}\left[\frac{\partial T^{\rm I}}{\partial E}\left(S^{\rm
  II}_{LR}{\cal N}^{\rm II}_{RL}S^{{\rm II}*}_{LL}+S^{\rm II}_{LL}{\cal N}^{\rm
  II}_{LR}S^{\rm II *}_{LR}\right)\right.\nonumber\\
&&\left.+\frac{\partial T^{\rm II}}{\partial E}\left(S^{\rm
  I}_{LR}{\cal N}^{\rm I}_{RL}S^{\rm I*}_{LL}+S^{\rm I}_{LL}{\cal N}^{\rm I}_{LR}S^{\rm I *}_{LR}\right)\right],
\end{eqnarray}
with $\delta\hat{n}^{\rm I/II}_R=\hat{n}^{\rm I/II}_R-\langle\hat{n}^{\rm I/II}_R\rangle$.
The cross-correlation depends on the off-diagonal elements of the density of
states matrix of Eq.\ (\ref{dens}). For two completely symmetric
scatterers with $S_{LR}=S_{RL}$ and $S_{RR}=S_{LL}$ it disappears.

{\begin{figure}[t]
\begin{center}
\psfrag{R} {$R$}
\psfrag{T} {$T$}
\psfrag{I} {${\rm I}$}
\psfrag{II} {${\rm II}$}
\psfrag{T1} {$T^{\rm I}$}
\psfrag{R1} {$R^{\rm I}$}
\includegraphics[width=5cm]{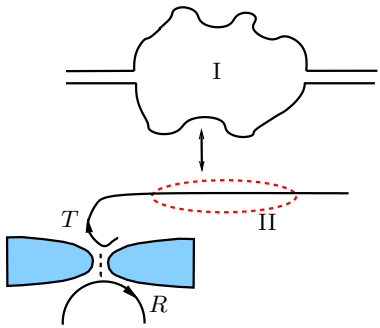}
\end{center}
\caption{Quantum dot (system ${\rm I}$) coupled to an edge state (system ${\rm
  II}$). The quantum point contact (with transmission probability $T$ and reflection
  probability $R$) is outside the interacting region.}
\label{edgestate}
\end{figure}}

Let us work out Eq.\ (\ref{corr1}) for two experimentally relevant systems. 
In a first example \cite{Ned07,NedPRL,examp1,examp2,examp3}
we assume that our second system is an edge state which is noisy
because of the presence of a
quantum point contact (QPC) left of it outside the interacting region,
as shown in Fig. \ref{edgestate} (this
is equivalent to assuming that the QPC has energy-independent transmission and reflection
probabilities $T$ and $R=1-T$). We do not specify
the first system. For an interacting region of length L the density of states is 
$N^{\rm II} = {m L}/{2\pi\hbar^2 k_F}$. 
We find
\begin{equation}
\label{corr2}
\langle \delta \hat{n}^{\rm I}_R\delta \hat{n}^{\rm II}_R \rangle = - \lambda
\, {\cal I}\,  \frac{\partial T^{\rm I}}{\partial E} \,{N^{\rm II}} \, {R T}.
\end{equation}
The factor $RT$ is a consequence of quantum partition of carriers at the QPC. 

With the experiment of Ref.\ \cite{McC07} in mind, we now assume that system II is a quantum dot with a Breit-Wigner
resonance at energy $E_F=E^{\rm II}$ (so ${\partial T^{\rm II}}/{\partial E}=0$ at $E_F$) and with
rates  $\gamma_L^{\rm II}$ ($\gamma_R^{\rm
  II}$) through the left (right) barriers. Conductor I is arbitrary. This gives
\begin{equation}
\label{corr3}
\langle \delta\hat{n}^{\rm I}_R\delta\hat{n}^{\rm II}_R\rangle=\lambda\, {\cal I}\frac{\partial T^{\rm I}}{\partial E}\, 
\frac{4\gamma_L^{\rm II}\gamma_R^{\rm II}(\gamma_L^{\rm II}-\gamma_R^{\rm II})}{\pi\left(\gamma_L^{\rm
  II}+\gamma_R^{\rm II}\right)^4}.
\end{equation}

Similar to the experiment \cite{McC07},  the correlations Eqs. (\ref{corr1}-\ref{corr3}) can have different signs.
The sign of the correlation depends on the asymmetry of the scattering matrix but also on the sign of the energy derivative
of the transmission probability \cite{Bro97}.

To conclude we have calculated the two-particle scattering matrix for two
weakly coupled mesoscopic conductors and we have expressed it as a function of the
non-interacting scattering matrices. We illustrated the properties of this matrix by calculating the transmission
probability and cross-correlation for a two-particle scattering experiment. Our results 
can be expressed in terms of the density of states matrices of the uncoupled conductors. 
The approach developed here is very general and permits to treat a large class of systems. 

We thank M. Polianski and D. S{\`a}nchez for their valuable comments.
The work was supported by the Swiss NSF and the EU Marie Curie RTN "Fundamentals of Nanoelectronics", MCRTN-CT-2003-504574.

\end{document}